\documentstyle[preprint,aps,epsf,floats]{revtex} 

\def\GeV{\ \rm{ GeV} }
\def\MeV{\ \rm{ MeV} }

\begin{document}

\preprint{\tighten \vbox{\hbox{
%CMU-HEP 99-01
} }}

\title{Bounds from Primordial Black Holes with a 
Near Critical Collapse Initial Mass Function}

\author{Graham D.\ Kribs, Adam K.\ Leibovich, and I.\ Z.\ Rothstein}

\address{
Department of Physics,
Carnegie Mellon University,
Pittsburgh, PA 15213-3890}

\maketitle

{\tighten
\begin{abstract}

Recent numerical evidence suggests that a mass spectrum of primordial 
black holes (PBHs) is produced as a consequence of near critical
gravitational collapse.  Assuming that these holes formed from the
initial density perturbations seeded by inflation, we calculate model
independent upper bounds on the mass variance at the reheating temperature 
by requiring the mass density
not exceed the critical density and the photon emission not exceed 
current diffuse gamma-ray measurements.  We then translate these results 
into bounds on the spectral index $n$ by utilizing the COBE data to 
normalize the mass variance at large scales, assuming a constant power
law, then scaling this result to the reheating temperature.
We find that 
our bounds on $n$ differ substantially ($\delta n>0.05$) from those 
calculated using initial mass functions derived under the assumption 
that the black hole mass is proportional to the horizon mass at the 
collapse epoch.  We also find a change in the shape of the diffuse 
gamma-ray spectrum which results from the Hawking radiation.
Finally, we study the impact of a nonzero cosmological constant
and find that the bounds on $n$ are strengthened considerably
if the universe is indeed vacuum-energy dominated today.

\end{abstract}
}%end tighten

%%\pacs{}

\newpage 

\section{Introduction}

Primordial black holes (PBHs) arise naturally in most cosmologies.
Perhaps the least speculative mechanism for PBH formation comes from
the collapse of overdense regions in the primordial density
fluctuations that gave rise to structure in the universe
\cite{carrhawking}.  Thus PBHs carry information of an epoch about
which we know comparatively little, and are a very useful tool for
restricting theories of the very early universe, especially within the
context of an inflationary scenario.  In 
this paradigm the density spectrum is  a consequence of the
quantum fluctuations of the inflaton field, and can in principle be
calculated given an underlying model.  Thus, the non-observation of
the by-products or of the effects of the energy density of these
PBHs constrains the underlying microscopic theory.

The simplest bound that can be extracted from PBH formation is generated 
by insisting that $\Omega_{PBH}\leq 1$.  Other bounds may be derived by
studying the consequences of their evaporation.  Given that black
holes evaporate at a rate proportional to their inverse mass 
squared~\cite{hawking},
the phenomenological relevance of a PBH will depend upon its initial
mass. Smaller mass PBHs ($10^9 < M_{bh} < 10^{13}$~g) will alter the
heavy elements abundances \cite{nuc} as well a distort the microwave
background \cite{micro}, whereas PBHs with larger masses will affect
the diffuse gamma-ray background \cite{diffuse,mac2,macreport}.

The net evaporation spectrum from a collection of PBHs will depend
on the initial mass distribution, which in turn depends upon the
probability distribution for the density fluctuations. 
Prior to the emergence of the COBE data, bounds from PBHs were
calculated assuming a Harrison-Zel'dovich spectrum $|\delta_k|^2 \propto k$, 
with an unknown normalization.  This lead to the 
famous ``Page-Hawking'' bound and all
its subsequent improvements \cite{diffuse,mac2}. 
However, assuming that the distribution is Gaussian, 
we can now relate the mass variance at the 
time of formation with the mass variance at large scales today, if 
we know the (model-dependent) power spectrum.
None the less, bounds on $n$ have been derived within the class of 
models where  the power spectrum is given by a power
law  $|\delta_k|^2 \propto k^n$ over the scales of
interest \cite{carrlidsey,cgl,green,kim1,kim2}. 
 Indeed, in this way it follows that for the 
scale-invariant Harrison-Zel'dovich spectrum PBHs have too small a number
density to be of any astrophysical significance.  However, observations (see
references in \cite{cgl}) now seem to favor a tilted blue spectrum
with $n>1$ (in CDM models) with more power at smaller scales, 
and thus the bounds derived from black hole evaporation can be used to
constrain the tilt of the
spectrum.

In this paper we revisit the aforementioned bounds in light of
recent calculations which indicate that a spectrum of primordial black 
hole masses are produced through near critical gravitational collapse. 
As was pointed
out by Jedamzik and Niemeyer \cite{JZI}, if PBH formation is a result
of a critical phenomena, then the initial mass function will be quite
different then what was expected from the classic calculation of Carr
\cite{carr}.  In particular, the PBH mass formed at a given epoch is no
longer necessarily proportional to the horizon mass.  The resulting 
difference in the initial mass function leads to new bounds, which is 
the main thrust of this paper.  In particular, we revisit the density bounds
and the bounds derived from the diffuse gamma-ray observations.
We will first derive bounds on $n$ in the class of models where
the power spectrum is a power law over the scales of interest.
We then relax this assumption and instead place
bounds on the mass variance at the formation epoch.\footnote{As will
be discussed later, PBH formation is dominated by the earliest
formation epoch if they form via critical collapse.} These bounds can
then be applied to a chosen model by extrapolating the
variance today to the formation epoch according to the appropriate
power spectrum.

Before continuing to the body of this work, we would like to point out
that primordial black holes have also played a large role in
attempting to explain various data. PBHs can serve as significant
cosmological flux sources for all particle species via Hawking
radiation\cite{hawking}. Thus it is very tempting to postulate that
present-day observed particle fluxes of unknown origin are a consequence 
of PBH evaporation.  However, to predict these fluxes, 
or model them realistically, we need to know the mass distribution of black
holes, since their emission spectra are determined by their
temperature (or inverse mass).  If we assume that the black holes of 
interest were formed from initial density inhomogeneities generated in 
an inflationary scenario (which is usually assumed), then the black holes 
are either tremendously over-abundant or completely negligible.  To get a 
phenomenologically interesting quantity of PBHs thus requires an extreme 
fine-tuning, as will be demonstrated below.  Succinctly, this fine-tuning 
arises because the PBH number density is an extremely rapidly varying 
function of the spectral index $n$.  Thus, without even analyzing the
details of the spectral profile, explaining unknown fluxes via PBH 
evaporation is far from compelling.

\section {The Initial Mass Function}

Carr \cite{carr} first calculated the PBH spectrum resulting from a 
scale-invariant Harrison-Zel'dovich spectrum up to an overall
normalization. Subsequently Page and Hawking calculated a bound on the
normalization by calculating the expected diffuse gamma-ray spectrum
from these PBHs \cite{diffuse,mac2}.  However, using the COBE measurements 
of the temperature anisotropies translated into density fluctuations  
(within a CDM), the overall normalization can be determined.
For a scale-invariant spectrum, no significant number density of PBHs
is generated.  However, for a tilted blue power spectrum
with more power on smaller scales, a larger number density
of PBHs is expected.

Given an initially overdense region with density contrast $\delta_i$
and radius $R$ at time $t_i$ (using the usual comoving coordinates),
analytic arguments predict \cite{carrhawking,carr} a black hole will
form if
\begin{equation}
1/3\leq \delta_i \leq 1 \; . 
\end{equation}
The lower bound on the density contrast comes from insisting that the
size of the region at the time of collapse be greater than the Jeans
length, while the upper bounds come from the consistency of the
initial data with the assumption of a connected topology.

The first calculations of the PBH mass distribution assumed the relation
\begin{equation}
\label{bhPropMh}
M_{bh}\simeq\gamma^{3/2}M_h,
\end{equation}
where $\gamma$ determines the equation of state $p = \gamma \rho$ and
$M_h$ is the horizon mass when the scale of interest crossed the
horizon.  Recently, numerical evidence suggests
that near the threshold of black hole formation, gravitational collapse
behaves as a critical phenomena with scaling and self-similarity
\cite{scaling}.  A scaling relation of the following form was found
\begin{equation}
\label{scaling}
M_{bh}(\delta)=kM_h(\delta-\delta_c)^{\rho},
\end{equation}
where $\rho$ is a universal scaling exponent which is independent 
of the initial shape of the density fluctuation.  It was
later shown \cite{JZI} that such scaling should be relevant for
PBH formation.  Indeed, in Ref.~\cite{JZII} the authors found relation
(\ref{scaling}) to hold for PBH formation when the initial conditions
are adjusted to be nearly critical. They found the exponent to be
$\rho\approx 0.37$. They also found that for several different initial
density shapes, $\delta_c\approx 0.7$, which is significantly larger
than the analytic prediction of $1/3$ found by requiring that the
initial overdensity be larger than the Jeans mass.

Given Eq.~(\ref{scaling}), calculating the initial PBH mass
distribution becomes analytically cumbersome, since in principle one
needs to sum over all epochs of PBH formation.  However, we would
expect that the initial mass function would be dominated by the
earliest epoch of formation if we assume a Gaussian distribution with
a blue power spectrum, since for larger scales the formation probability
should be suppressed.  This expectation was tested in
Ref.~\cite{green2}, where the authors used the excursion set formalism
\cite{bond} to calculate the initial mass function allowing for PBH
formation at all epochs. The authors found that it was approximately true
 the
 that earliest epoch dominates, for the conditions
of interest to us.

This simplification allows us to derive quite easily the initial mass
distribution \cite{JZI}.  We assume Gaussian fluctuations (the effects
of non-Gaussianity will be briefly discussed in Sec.~\ref{robust-sec}) 
and define the usual smoothed density contrast
\begin{equation}
\delta_R(x)=\int d^3y\,\delta (x+y) W_R(y) \; ,
\end{equation}
where $\delta(x)=(\rho(x)-\rho_b)/\rho_b$, and $\rho_b$ is background
energy density. $W_R$ is the window function with support in a region
of size $R$. The probability that a region of size $R$ has density
contrast between $\delta$ and $\delta+d\delta$ is given by
\begin{equation}
P(R,\delta)d\delta =\frac{1}{\sqrt{2 \pi}\sigma_R}\exp\left(-\frac{\delta^2}
{2\sigma_R^2}\right) d\delta \; ,
\end{equation}
where $\sigma_R$ is the mass variance for a region of size $R$, 
$\sigma^2_R=\langle \delta_R^2(x) \rangle/R^3$.

Then using Eq.~(\ref{scaling}), the physical number density of PBHs
within the horizon per logarithmic mass interval, at the formation
epoch, can be written as
\begin{equation}
\label{IMF1}
\frac{d n_{bh}}{d\log M_{bh}} = 
   V_h^{-1}\,P[\delta(M_{bh})]\,\frac{d\delta}{d\log M_{bh}} = 
   \frac{V_h^{-1}}{\sqrt{2 \pi}\,\sigma\,\rho} 
   \left( \frac{M_{bh}}{k M_H} \right)^{1/\rho}\,
   \exp\left[-\frac{1}{2\sigma^2} \left[ \delta_c 
             + \left(\frac{M_{bh}}{k M_H} \right)^{1/\rho} \right]^2 
   \right], 
\end{equation} 
where $V_{h}$ is the horizon volume the the epoch of PBH formation.
We assume prompt reheating, and therefore take the formation epoch to
be the time of reheating, which corresponds to the minimum horizon
mass~\cite{green2}.

Assuming that the power law spectrum holds down to the
scales of the reheat temperature, then $\sigma_H^2 \propto R^{-(n+3)}$.
We can then relate the mass variance today $\sigma_0$ to the mass
variance at the epoch of PBH formation $\sigma(M_H)$, using 
\cite{green,green2}
\begin{equation}
\label{dis1}
\sigma^2(M_H)=\sigma_0^2 \left( \frac{M_{eq}}{M_0} \right)^{(1-n)/3}
\left( \frac{M_{H}}{M_{eq}} \right)^{(1-n)/2},
\end{equation}
where
\begin{equation}
M_H = M_0\left(\frac{T_{eq}}{T_{RH}}\right)^2
   \left(\frac{T_0}{T_{eq}}\right)^{3/2}, 
\end{equation}
$M_0$ is the mass inside the horizon today, $T_{RH}$ is the reheat
temperature, and $T_{eq}$ is the temperature at radiation-matter equality.
This relation is essential to connect present-day density fluctuations
to those of much earlier times.  Two important assumptions underly
this useful relation:  First, $n$ is taken to be constant over all
the scales of interest.  Second, the universe was assumed to be
radiation dominated until the temperature dropped below 
$T_{eq} \sim 5$~eV (matter-radiation equality), and then matter
dominated thereafter.

From the COBE anisotropy data, the mass variance can be calculated
\cite{bunn,green}
\begin{equation}
\label{COBE}
\sigma_0=9.5 \times 10^{-5} .
\end{equation}
Using this result we can then calculate the physical number density 
per unit mass interval at $T = T_{RH}$
\begin{equation}
\label{IMF}
\frac{d n_{bh}}{dM_{bh}}= 
  \frac{V_h^{-1}}{\sqrt{2\pi}\sigma(M_H)M_{bh}\rho}y^{1/\rho}\exp\left[
  -\frac{\left(\delta_c+y^{1/\rho}\right)^2}{2\sigma^2(M_H)}\right],
\label{phys-num-density-eq}
\end{equation}
where
\begin{equation}
y=\frac{M_{bh}}{kM_H}=\frac{M_{bh}T_{RH}^2}
{0.301 k g_\star^{-1/2} M_{Pl}^3} \; ,
\end{equation}
and $M_{Pl}$ is the Planck mass.  The physical number density 
at time $t$ is simply Eq.~(\ref{phys-num-density-eq}) rescaled
by a ratio of scale factors, that can be written as
\begin{equation}
\frac{d n_{bh}}{d M_{bh}}(t) = \left( \frac{T(t)}{T_{RH}} \right)^3 
                               \frac{d n_{bh}}{d M_{bh}} \; .
\label{scaled-density-eq}
\end{equation}

Finally, we should note that relation (\ref{scaling}) is only valid
for $\delta \approx \delta_c$. Thus we expect that we may integrate
over $\delta$ with small errors, as long as the width of the Gaussian
is sufficiently small. In particular, our results should be trustworthy
provided $\sigma \lesssim 1$, which implies $n$ should not exceed 
the maximum
\begin{equation}
n_{max}\simeq 1+\frac{2 \log(\sigma_0^2)}{\log(T_{eq}T_0/T_{RH}^2)} \; .
\end{equation}
We will see that $n$ does not exceed this maximum value for
all of the bounds we consider.

\section{Bounds from $\Omega \leq 1$}

Let us now calculate the total energy density in PBHs.  We assume the
``standard cosmology'' where the universe began in an inflationary
phase, reheated, was radiation dominated from the reheating period 
until matter-radiation equality, and then has been matter dominated.
The contribution of a PBH with a given initial mass, $M_{bh}$, 
to the energy density today will depend upon its lifetime. 
The time-dependent PBH mass $M(t)$ is given 
by~\cite{hawking}
\begin{equation}
\label{massevol}
M(t) = M_\star\left[\left(\frac{M_{bh}}{M_\star}\right)^3
   -\frac{t}{t_0}\right]^{1/3},
\end{equation}
where $M_\star$ is the initial mass of a PBH which would be decaying 
today, $M_\star \simeq 5 \times 10^{14}$~g.  It is a good approximation 
to assume that the black hole decays instantaneously at a fixed decay 
time, $t_d$, which we use in the following.

There are two components to the PBH density bounds that we can
calculate.  The first is the total energy density of the PBHs that
have not decayed by a given time $t$.  The second is the total
energy density of the products of PBH evaporation.  The sum of these
components must be less than the critical density
$\Omega_{pbh,evap} + \Omega_{pbh} < 1$, at any time.
The evaporated products of PBHs, in particular photons, 
could break up elements during nucleosynthesis, disrupting the 
well-measured elemental abundances.  This and other processes 
during nucleosynthesis provide additional bounds on the density of
PBHs~\cite{nuc} that we do not discuss here.

The simplest bound comes from the density of PBHs that have not
decayed by time $t$, 
\begin{equation}
\rho_{pbh}(t) = \left( \frac{T(t)}{T_{RH}} \right)^3 
                \rho_{tot}(t_{RH}) \> I_{M_\star(t)}^{M_{max}}(0)
\end{equation}
where $T(t)$ is the temperature of the universe at time $t$, and $M_\star(t)$ 
is the initial PBH mass that has just completely evaporated by time $t$,
\begin{equation}
M_\star(t) \approx M_\star \left( \frac{t}{t_0} \right)^{1/3} \; .
\end{equation}
$I_{M_1}^{M_2}(\xi)$ is a dimensionless weighted integral over the 
PBH mass spectrum between $M_1$ to $M_2$,
normalized to the total density $\rho_{tot}(t_{RH})$,
\begin{equation}
I_{M_1}^{M_2}(\xi) = \frac{1}{\rho_{tot}(t_{RH})} 
                     \int_{M_1}^{M_2} d M_{bh} M_{bh} 
                     \frac{d n_{bh}}{d M_{bh}} 
                     \left( \frac{M_{bh}}{M_\star} \right)^\xi \; ,
\end{equation}
where $d n_{bh}/d M_{bh}$ is given by Eq.~(\ref{phys-num-density-eq}).
We can use the above to trivially compute the ratio of the PBH density to
the critical density, $\Omega(t)$.  In particular, we need 
only compute the density ratio at three relevant epochs: immediately after 
reheating $t=t_{RH}$, at matter-radiation equality $t=t_{eq}$, and 
present-day $t=t_0$.  The density ratios are\footnote{Note that
$\Omega(t_{RH}) = I_0^{M_{max}}(0)$ is often denoted by $\beta(t_{RH})$.}
\begin{eqnarray}
\Omega(t_{RH}) &=& I_0^{M_{max}}(0) \\
\Omega(t_{eq}) &=& \frac{T_{RH}}{T_{eq}} 
                   I_{M_{eq}}^{M_{max}}(0) \\
\Omega(t_{0})  &=& \frac{T_{RH}}{T_{eq}} 
                   I_{M_\star}^{M_{max}}(0) \; .
\end{eqnarray}
where $M_{max}$ corresponds to $\delta = 1$, which is of order the
horizon mass at reheating.  The integral should be independent of the 
upper limit $M_{max}$ if we are to trust our results.

Making the conservative approximation that all the PBH decay
products are relativistic, the contribution to the density ratio of
the products of PBH evaporation that has occurred up until today can
be written as
\begin{equation}
\Omega_{pbh,evap}(t_0) = \frac{T_{RH}}{T_{eq}} \left( \frac{T_0}{T_{eq}} 
    \right)^{1/4} I_{0}^{M_{eq}}(3/2) +
    \frac{T_{RH}}{T_{eq}} I_{M_{eq}}^{M_\star}(2) \; .
\end{equation}
In Fig.~\ref{bound-fig}, we show the upper limit on $n$ as a function
of $T_{RH}$ coming from bounding $\Omega_{pbh,evap}(t_0) +
\Omega_{pbh}(t_0) < 1$ (solid line).  For larger values of the reheat
temperature we get a more stringent bound by imposing the constraint
$\Omega_{pbh}(t_{RH}) \leq 1$, simply because as $T_{RH}$ is increased
more of the black holes will have decayed at an earlier epoch.  Given
that most of the energy of the decay products resides in radiation,
the effect on $\Omega_{pbh}(t_0)$ is diminished due to the redshifting.
This new bound is given by the dotted line in Fig.~\ref{bound-fig}.

If we assume that the PBH leaves behind a Planck mass remnant, then we
have additional bounds which become important for very large reheat
temperature\cite{mac,barrow,cgl}.  The best bound in this case comes 
from calculating $\Omega_{remnant}(t_{eq})$ which is given by
\begin{equation}
\Omega_{remnant}(t_{eq}) = \frac{T_{RH}}{T_{eq}} \frac{M_{Pl}}{M_\star} 
                           I_0^{M_{eq}}(-1) \; .
%\Omega_{remnant}(t_{eq}) = \frac{T_{RH}}{T_{eq}} 
%\int_0^{M_{eq}} M_{Pl}
%\frac{d n_{bh}}{dM_{bh}}d M_{bh}.
\end{equation}
The bound in this case is shown as the dashed line in Fig.~\ref{bound-fig}, 
and is the best bound at the largest values of the reheat temperature.

\section{Bounds from Diffuse Gamma-Rays}

For a certain range of $T_{RH}$ we can improve our bounds on $n$ from
diffuse gamma-ray constraints.  The present day flux is determined by
convoluting the initial mass function with the black hole emission
spectrum
\begin{equation}
\label{hawkingspectrum}
f(x)=\frac1{2\pi}\frac{\Gamma_s(x)}
{\exp(8\pi x)-(-1)^{2 s}},
\end{equation}
where $s$ is the spin of the emitted particle, $x=\omega(t) M(t)/M_{Pl}^2$, 
$\omega(t)$ is the frequency and $M(t)$ is the PBH mass at the time $t$
of emission.  $\Gamma_s(x)$ is the absorption coefficient and may be 
written as $[\omega(t)]^2 \sigma_s/\pi$.  
$\sigma_s$ is the absorption cross section and is calculated using the
principle of detailed balance.

The values for $\sigma_s$ were calculated some time ago by
Page\cite{page1,page2}.  Let us consider how $\sigma_s$
behaves for massless particles. At large values of $x$, $\sigma_s$
performs small oscillations about the geometric optics limit of
$\sigma_g=27 \pi M^2/M_{Pl}^4$. As $x$ approaches zero, $\sigma_s$
goes to zero for $s=1/2,1$ but goes to a constant value for $s=0$.  We
will use the approximation
\begin{equation}
\Gamma_s(x)= (56.7,~20.4) x^2/\pi~~ $\rm for$~~ s=(\frac12,~1).
\end{equation}
This approximation is poor at low energies, as it is in
error by $50\%$ at $x=0.05$.  However, as we shall see, the
contribution to the spectrum of interest is greatly peaked at
$x\simeq0.2$.  The case of strongly interacting particles is
complicated by the hadronization process. There is a large
contribution coming from pion decay, however, given the extreme
sensitivity of the flux to the value $n$, the effect on the bound is
negligible.

It has been recently suggested \cite{heckler1,heckler2} that the 
self-interactions of the emitted particles will induce a photosphere, thus
distorting the spectrum considerably from Eq.~(\ref{hawkingspectrum}).
It was suggested that two types of photospheres should form. A QCD
photosphere\footnote{In the case of QCD what is meant by
``photosphere'' is a quark-gluon cloud.}  generated by parton-parton
interactions as well as a QED photosphere generated by
electron-positron-photon interactions.  This is idea has been
tested more quantitatively via a numerical solution of the Boltzmann
equation~\cite{mcgill}. 
Again, while this effect may change the spectrum, especially
at higher energies, it is irrelevant as far as the bound 
on the spectral index is concerned.

The flux measured today is given by
\begin{equation}
\label{cosmflux}
  \frac{dJ}{d\omega_0} = \frac{1}{4\pi} \int_{t_i}^{t_0} dt (1+z) 
  \int d M_{bh} \, \frac{d n_{bh}}{dM_{bh}}(t) \, f(x) \; ,
\end{equation}
where $d n_{bh}/d M_{bh}$ is evaluated at time $t$ using 
Eq.~(\ref{scaled-density-eq}), $t_0$ is the age of the universe, 
$t_i$ is the time of last scatter, and $f(x)$ is the instantaneous 
emission spectrum given above with
\begin{equation}
x = \frac{\omega(t) M(t)}{M_{Pl}^2} = \frac{\omega_0 (1+z)}{M_{Pl}^2}
  M_\star \left[ \left( \frac{M_{bh}}{M_\star} \right)^3 
                 - \frac{t}{t_0} \right]^{1/3} .
\end{equation}
The integral over $t$ is cut off at early times, since at redshifts
above $z = z_0 \simeq 700$ the optical depth will be larger than unity
due to either pair production off of matter or ionized matter
\cite{ZS}.  Those processes will degrade the energy below the window
we are interested in.

This integral may be rewritten in the more illuminating form
\begin{equation}
\label{cosmoSpectrum}
  \frac{dJ}{d\omega_0} =
  \frac{1}{4\pi}\frac{M_{Pl}^6}{(\omega_0M_\star)^3} 
  \int_{0}^{z_0}\frac{dz}{ H_0 (1+z)^{5/2}}
  \int_{0}^{\infty} dx \> x^2 \alpha^{-2}\,f(x)\frac{d n_{bh}(x,z)}{dM_{bh}},
\end{equation}
where
\begin{equation} 
\label{mbh}
\alpha=\frac{M(t)}{M_\star}=\left\{(1+z)^{-3/2} +
	\frac{x^3M_{Pl}^6}{[(1+z)\omega_0M_\star]^3}\right\}^{1/3}.
\end{equation}

\begin{figure}[!t]
\centering
\epsfxsize=6.4in
\hspace*{0in}
\epsffile{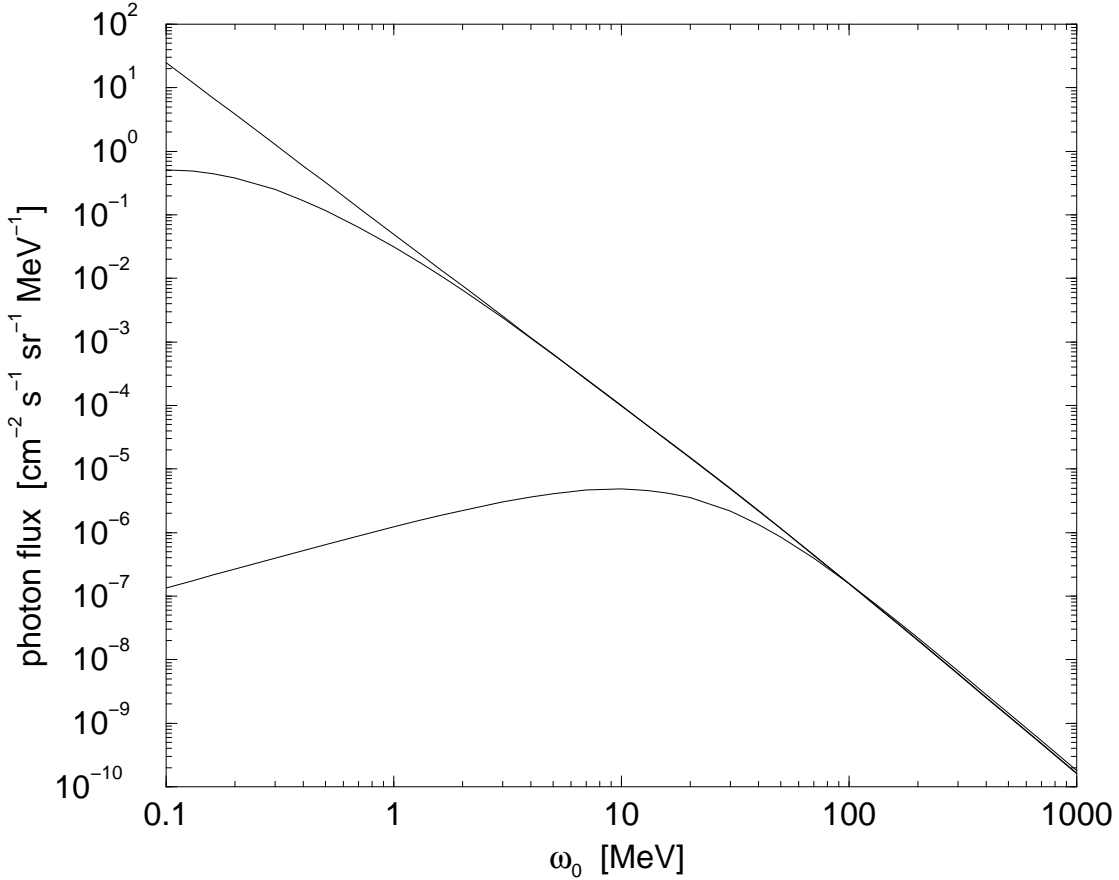}
\caption{Examples of the direct cosmological photon spectrum from
evaporating PBHs with $n \approx (1.340, 1.309, 1.285)$,
$T_{RH}=(10^7, 10^8, 10^9)$~GeV for the (top, middle, bottom) solid
lines.  Notice that the spectrum scales roughly as $dJ/d\omega_0
\propto \omega_0^{-3}$ for $\omega_0 > \omega_{kink}$.  At low
energies ($\omega_0 \ll 100\MeV$), the spectrum is modified by the
production of quarks and gluons emitted from the PBHs that fragment
into pions, which then decay into photons.  At high energies
($\omega_0 \protect\gtrsim 100\MeV$), the spectrum is modified if a
photosphere forms around the PBH.  Since the normalization of the PBH
mass spectrum is extremely sensitive to the spectral index, both
these effects are completely negligible when calculating the bound on
$n$.}
\label{examples-fig}
\end{figure}

Let us study the qualitative behavior of the above integral as a
function of $\omega$ at fixed $n$ and $T_{RH}$. The $x$ integration is
controlled by the Boltzmann factor in $f(x)$. Indeed, a little
manipulation shows the the integrand is highly peaked near
$x \simeq 0.2$. Furthermore, the $\omega$ dependence in  $\alpha^{-2}$
is almost completely canceled by the $\omega$ dependence in the
factor $M_{bh}^{-1}y^{1/\rho}\propto\alpha^{(1/\rho-1)}$ in
$d n_{bh}/dM_{bh}$. Thus the $\omega$ dependent part of the integrand may be
written as
\begin{equation}
\frac{dJ}{d\omega_0}\propto \omega_0^{-3} 
\exp\left[-\frac{\left(\delta_c+aT_{RH}^{2/\rho}\alpha^{1/\rho}\right)^2}
                {2\sigma^2(M_H)}\right],
\end{equation}
where $a^\rho = M_\star g_\star^{1/2}/(0.301 k M_{Pl}^3)$, and the
only $\omega$ dependence in the exponential is through $\alpha$.  If
for now we assume that the dominant contribution the higher energy
photons comes from recent decays ($z\sim 0$), and most of the support
for the $x$ integral comes with $x \sim 0.2$, then $\alpha$ simplifies to
\begin{equation}
\alpha \approx \left[ 1 + \left( \frac{0.2 M_{Pl}^2}{w_0 M_\star} \right)^3 
               \right]^{1/3}.
\end{equation}
As $\omega_0$ gets larger than $0.2 M_{Pl}^2/M_\star \sim 100\MeV$,
$\alpha$ becomes independent of $\omega_0$ and therefore the flux
behaves as $dJ/d\omega_0 \propto \omega_0^{-3}$.  
For lower energies we can make the
approximation $\alpha\sim 0.2 M_{Pl}^2/(w_0M_\star)$, and we would
expect that at some point the $\omega$ dependence in the exponential
will begin to dominate such that the flux should begin to rapidly decrease
as we go to lower photon energies.  The energy at which the flux turns
over is determined by the competition between the two terms $\delta_c$ and
$a T_{RH}^{2/\rho}\alpha^{1/\rho}$ in the exponential, which is set by
the reheat temperature.  As we lower the reheat temperature the
position of the kink moves to lower energies.  If the reheat
temperature is higher than $T_{RH}\sim 10^9\GeV$, however, the peak
will stay around 100 MeV, since at these temperatures the second term
in the exponential will always dominate.  Indeed, we expect the
position of the fall off to be near
\begin{equation}
\label{kink}
\omega_{\rm kink}\simeq \min\left(100{\MeV}\,,\,
  \frac{0.2g_\star^{1/2}\,T_{RH}^2}
    {0.301\,k\,M_{Pl}\,\delta_c^{\rho}}\right).
\end{equation}
Figure~\ref{examples-fig} shows the flux for fixed $n$ for a few 
different reheat
temperatures.  The position of the kink is well tracked by
Eq.~(\ref{kink}).  Note however that the flux does not fall off
exponentially at energies below the kink.  This is because as we go to
lower energies we pick up more of a contribution from higher
redshifts.

It is interesting to contrast this behavior with the flux calculated
assuming that the mass of a PBH formed at a given epoch is proportional to
the horizon mass at the time of collapse.  In Refs.~\cite{kim1,kim2} the
authors calculated an initial mass function following the
Press-Schecter formalism, summing over all epochs and assuming the
relation $M_{bh}\simeq\gamma^{3/2} M_H$ at each epoch. They found
\begin{equation}
\label{KLM}
\frac{d n_{bh}}{dM_{bh}} = \frac{n+3}{4} \sqrt{\frac{2}{\pi}} \gamma^{7/4} 
  \rho_i M^{1/2}_{H_i}M_{bh}^{-5/2}\sigma_H^{-1} 
  \exp\left(-\frac{\gamma^2}{2\sigma_H^2}\right),
\end{equation}
where $\rho_i$ and $M_{H_i}$ are the energy density and horizon mass
at $T_{RH}$ and
\begin{equation}
\sigma_H=\sigma_0\left(\frac{M_{bh}}{\gamma^{3/2}M_0}\right)^{(1-n)/4}.
\end{equation}
This result reduces to  the initial mass function first computed by
Page \cite{carr} for the Harrison-Zel'dovich spectrum with $n=1$ and
$d n_{bh}/dM_{bh} \propto M_{bh}^{-5/2}$.  The $\omega_0$ dependence
of this result arises only through $M(t)$ given by Eq.~(\ref{mbh}).
Using this initial mass distribution in Eq.~(\ref{cosmflux}), we
expect, as in the previous case, $dJ/d\omega_0 \propto \omega_0^{-3}$
for larger energies, and exponential decay into the lower energies
(which will again be mollified from photons descending from higher
redshifts).  However, for this case the position of the kink will be
fixed at around 100 MeV, independent of the reheat temperature.

Let us compare the above prediction with the recent COMPTEL and EGRET
data. The EGRET collaboration found that the flux in the energy range
$30\MeV-100\GeV$ is well fit by the single power law \cite{EGRET}
\begin{equation}
\label{EGRET}
\frac{dJ}{d\omega_0}=7.32\pm 0.34\times 10^{-9}
   \left(\frac{\omega_0}{451\MeV}\right)^{-2.10\pm 0.03}~
   ({\rm cm^2~sec~sr~MeV})^{-1},
\end{equation}
while the COMPTEL data \cite{COMPTEL} in the range $0.8$--$30\MeV$ 
can be fit \cite{kribs} to the power law
\begin{equation}
\frac{dJ}{d\omega_0}=6.40\times 10^{-3}
   \left(\frac{\omega_0}{1\MeV}\right)^{-2.38}~
   ({\rm cm^2~sec~sr~MeV})^{-1}.
\end{equation} 
Below $0.8\MeV$ there is large increase in the measured flux.  Thus,
the best bounds are found by comparing the measured flux to predicted
flux at $\omega_{\rm kink}$ or at $0.8\MeV$, whichever is
larger. Because of the rapid rise of the predicted spectrum relative
to the measured spectrum, a change in the kink position can change the
bound on $n$ on the order of $0.01$, which we consider within the
accuracy of our calculation.  The bounds on $n$ from the diffuse
gamma-rays are specified by the dot-dashed line in
Fig.~\ref{bound-fig}.  The bound terminates when all but the
exponential tail of the PBHs decay prior to a redshift of $700$, since
the optical depth at such early times exceeds unity, as discussed
above.

We may compare our results to those derived by Yokoyama
\cite{yokoyama},\footnote{After this work was completed, we
became aware of Ref.~\cite{greensusy} that also utilized the
critical collapse initial mass function to derive bounds on the
PBH mass density by requiring that LSPs (in supersymmetric models) 
not be overproduced.} where the author placed bounds on mass fraction 
of PBHs at $t_{RH}$, $\beta(t_{RH}) = \Omega(t_{RH})$, using the 
initial mass function, Eq.~(\ref{IMF}).  
He found that the bounds on $\beta$ did not differ
significantly from the previous bounds derived using the standard
initial mass functions, except for the bounds coming from diffuse
gamma-rays.  In the latter case, applicable for horizon masses in the
range $M_H\geq 5\times 10^{14}$~g, he found more stringent
constraints. Our bounds on $n$, translated into bounds on $\beta$,
agree with his bounds coming from energy density constraints except
for the case of larger reheat temperature, since we included the proper
scaling of the energy density of photons emitted after PBH decay.  Thus our
bounds on $\beta$ can differ by many orders of magnitude.  Our bounds
coming from diffuse gamma-rays can also differ by orders of magnitude,
but in this case it is for a different reason. Yokoyama determined
his bound on $\beta$ by imposing the constraint on $\Omega_{pbh}(t_0)$
derived in Ref.~\cite{mac2}. However, when we change the initial mass
function we also change the diffuse gamma-ray spectrum significantly
in both shape and normalization, as discussed above.  Thus, it is
inappropriate to directly take the bounds from Ref.~\cite{mac2} and
apply them to the case with the new initial mass function,
Eq.~(\ref{IMF}). We find that our bounds on $\beta$ from diffuse
gamma-rays are more stringent than those determined in
Ref.~\cite{yokoyama} in the range $M_H>5 \times 10^{15}$~g by several
orders of magnitude.

\section{Robustness of the  Bounds}
\label{robust-sec}

Let us consider the robustness of the bounds. We might worry that the
bounds are highly sensitive to the choice of parameters given the
sharpness of the initial mass function. Indeed, in the case where it
is assumed that the PBH mass is given by Eq.~(\ref{bhPropMh}), the
bounds are $n$ are exceptionally sensitive to the exactness of this
relation. This is clear from the exponential factor in
Eq.~(\ref{KLM}).  Given the initial mass function calculated
by Jedamzik and Niemeyer, Eq.~(\ref{IMF}), 
we must check the sensitivity to the parameters $\delta_c,~k$ and
$\sigma_0$. In Ref.~\cite{JZII}, the authors tested the scaling relation
(\ref{scaling}) using several different initial shapes density
perturbations shapes. They found $(\delta_c=0.70, ~k=11.9)$,
$(\delta_c=0.67, ~k=2.85)$, $(\delta_c=0.71, ~k=2.39)$, for Gaussian,
Mexican Hat and fourth order polynomial fluctuations, respectively.

We varied the value of $\delta_c$ between $0.60-0.80$ and found that
the bounds changed by at most $0.01$.  The sensitivity to the
variation being maximal at the smaller values of $T_{RH}$. Given that
the initial mass function is peaked at a number smaller than $k M_h$, the
sensitivity is increased at smaller $T_{RH}$ because $\sigma$ is an
decreasing function of $T_{RH}$.  Variations in $k$ are equivalent to
a scaling in $T_{RH}$. Thus, varying $k$ by an order of magnitude has
essentially no effect on the bound. Lastly, let us consider the
sensitivity to the parameter $\sigma_0$. The value we used for
$\sigma_0$ in Eq.~(\ref{COBE}) was calculated in Ref.~\cite{green2} 
using the result \cite{bunn}\footnote{It should be emphasized 
that this result assumed the spectra
can be approximated as a power law over the range of $k$ that COBE
probes.  We are then making the further assumption that $n$ is
constant down to the mass scales of relevance for PBHs.}
\begin{equation}
\label{fit}
\delta_0=1.91\times 10^{-5} \frac{\exp[1.01(1-n)]}{\sqrt{1+0.75 r}} \; ,
\end{equation}
where $r$ is a measure of the size of the tensor perturbations.  The
$1\sigma$ observational error being $7\%$. The fit, Eq.~(\ref{fit}),
is good to within $1.5\%$ everywhere within the region $0.7 \leq n
\leq 1.3$ and $0\leq r \leq 2$. The authors of Ref.~\cite{bunn} quote
a $9\%$ uncertainty in Eq.~(\ref{fit}) at $1\sigma$, once
uncertainties in the systematics and variations in the cosmological
parameters are taken into account.  The value in Eq.~(\ref{COBE}) was
determined ignoring tensor perturbations.
%, and taking $n\simeq 1.2$. 
Given that $\sigma_0$ scales with $\delta_0$ we find that
varying $\sigma_0$ at the $2\sigma$ level has no effect on our bound
at the level of $0.01$. On the other hand, including some contribution
from tensor perturbation will weaken the bound. We found that taking
$r=2$ weakened the bound by $0.01-0.02$ throughout the range in the
reheat temperature.  

We can also consider the effects of a non-vanishing $\Omega_\Lambda$.  
Bunn et~al.~\cite{bunn} extended their results to this case and found
\begin{equation}
\label{labmda}
\delta_0|_{\Omega_\Lambda}=1.91\times 10^{-5} \frac{\exp[1.01(1-n)]}
{\sqrt{1+(0.75-0.13\,\Omega_\Lambda^2)r}}\Omega_0^{-0.80-0.05\log{\Omega_0}}
\, \left(1+0.18\,(n-1)-0.03\,r\,\Omega_\Lambda \right) \> .
\end{equation}
If $0 \le r \le 2$, we can express $\delta_0|_{\Omega_\Lambda}$ 
extracted assuming a nonzero cosmological constant to a very good 
approximation by a scaling of $\delta_0$ extracted without a 
cosmological constant
\begin{equation}
\delta_0|_{\Omega_\Lambda} \approx
    \Omega_0^{-0.80-0.05\log{\Omega_0}} \> \delta_0 \; ,
\end{equation}
(where $\Omega_0 + \Omega_\Lambda = 1$)
and thus $\sigma_0$ also acquires a correction.  Consequently,
the bound on $n$ is shifted by
\begin{equation}
\Delta n \equiv n - n|_{\Omega_\Lambda} = 
    \frac{2 \> (-0.8 - 0.05 \ln \Omega_0) \, \ln \Omega_0}{42.9 
    + \ln (T_{RH}/10^8 \; \mathrm{GeV})} \; .
\end{equation}
In Fig.~\ref{delta-n-fig} we show the above correction as a function 
of $T_{RH}$ for several choices of $\Omega_\Lambda$.  If we take
$\Omega_\Lambda \approx 0.7$ as recent observational data suggests,
our bounds on $n$ strengthen by about $0.03-0.06$ for $T_{RH}$ between
$10^{16}-10^{3}$ GeV respectively, as shown in Fig.~\ref{bound-lambda-fig}.

We can also calculate bounds on the mass variance at reheating 
\cite{liddle-grqc} which is essentially model-independent.  
If the relation Eq.~(\ref{dis1}) is violated by, for example, a power 
spectrum with a spectral index that depends on scale, then our previous
bounds on $n$ cannot be applied.  However, given a inflationary
model one could in principle calculate the power spectrum, 
normalize to the COBE data at our present epoch, 
and then match onto the mass variance at reheating.  
In Fig.~\ref{sigma-fig} we show the bounds on
the mass variance from both the density bounds as well as the
bounds from the diffuse gamma-ray observations.  Notice that
the diffuse gamma-ray observation bounds on $\sigma(M_H)$
are a significant improvement over the density bounds in the
applicable range of reheating temperatures.

Finally, we must address the issue of non-Gaussianity. It has been
pointed out \cite{bullock} that skewness could very well be important
for PBH formation given that its effects are amplified in the tail of
the distribution $P[\delta]$, which contributes to PBH formation.  In
general, the amount of non-Gaussianity expected is highly model
dependent. Bullock and Primack investigated several inflationary
models to study the amount of non-Gaussianity one would expect at
larger values of $\delta$. They calculated $P[\delta]$ for three toy
models, and found in one case no deviation from Gaussianity and in the
other two found a significant suppression in the probability of of
large perturbations. However, as was pointed out in Ref.~\cite{green},
while these effects can drastically effect the PBH mass fraction
$\beta$, we expect that, even in the most extreme case considered in
Ref.~\cite{bullock}, the effect on the bound on $n$ is only at the
level of $0.05$.  For hybrid inflation, where the approximation that
$n$ is constant actually holds, the perturbations are in fact
Gaussian due to the linear dynamics of the inflaton field \cite{Yi}.
Therefore, these bounds should be applied to specific models, with the
roughness of the bound determined by the deviations away from
Gaussianity.

\section{Conclusions}

\begin{figure}[!t]
\centering
\epsfxsize=6.4in
\hspace*{0in}
\epsffile{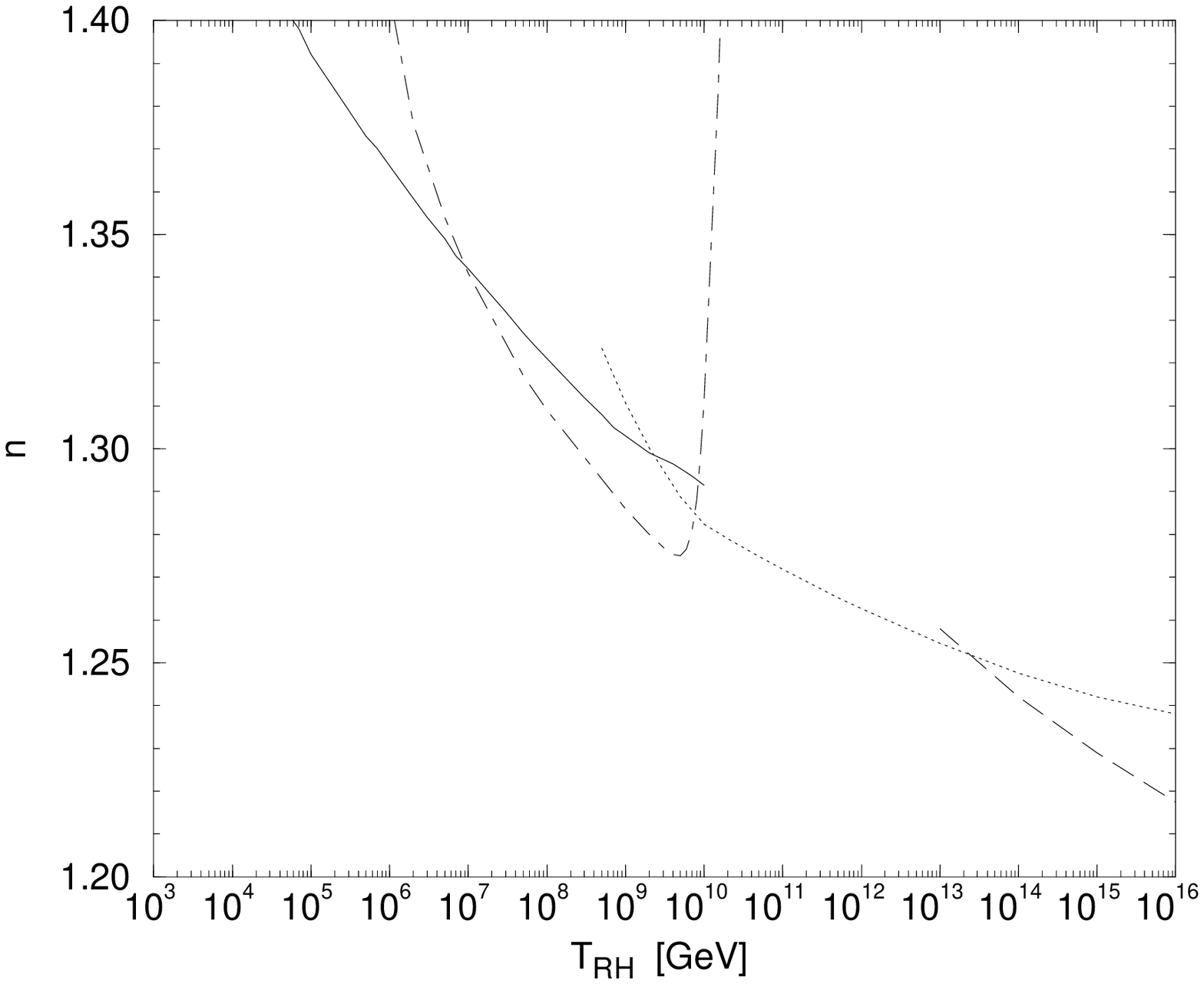}
\caption{The upper bound $n$ as a function of $T_{RH}$ by requiring
that $\Omega_{pbh,evap}(t_0) + \Omega_{pbh}(t_0) < 1$ (solid line),
$\Omega_{pbh}(t_{RH}) < 1$ (dotted line), and that the PBH photon spectrum 
does not exceed the diffuse gamma-ray background (dot-dashed line).
If PBHs leave a Planck mass relic, an additional bound is present
for large $T_{RH}$ (dashed line).  Note that the endpoint of each line 
within the figure is where we elected to stop calculcating the bound on $n$, 
due to the presence of another stronger constraint as shown.}
\label{bound-fig}
\end{figure}

\begin{figure}[!t]
\centering
\epsfxsize=6.4in
\hspace*{0in}
\epsffile{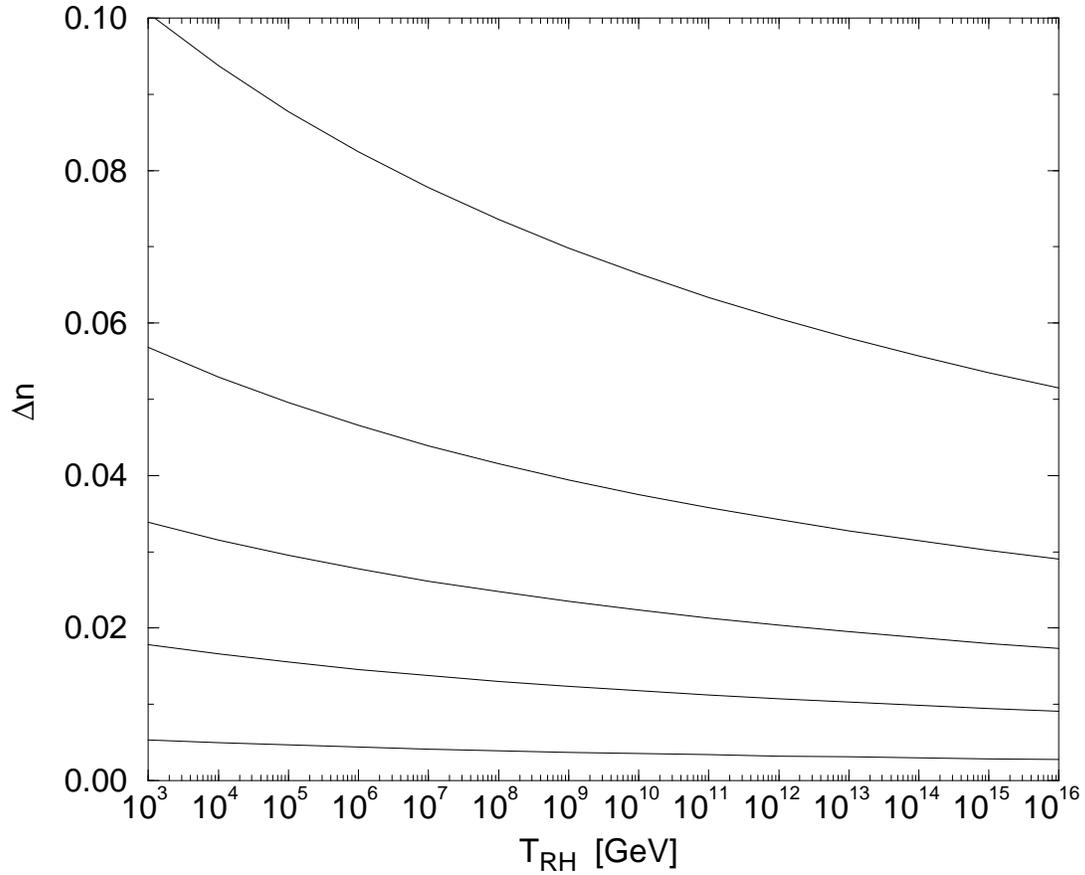}
\caption{The correction 
$\Delta n \equiv n - n|_{\Omega_\Lambda}$ as a function of $T_{RH}$
to all our bounds in Fig.~\ref{bound-fig} induced by assuming a 
cosmological constant $\Omega_\Lambda = (0.9, 0.7, 0.5, 0.3, 0.1)$ 
for the five lines in the figure from top to bottom respectively.}
\label{delta-n-fig}
\end{figure}

\begin{figure}[!t]
\centering
\epsfxsize=6.4in
\hspace*{0in}
\epsffile{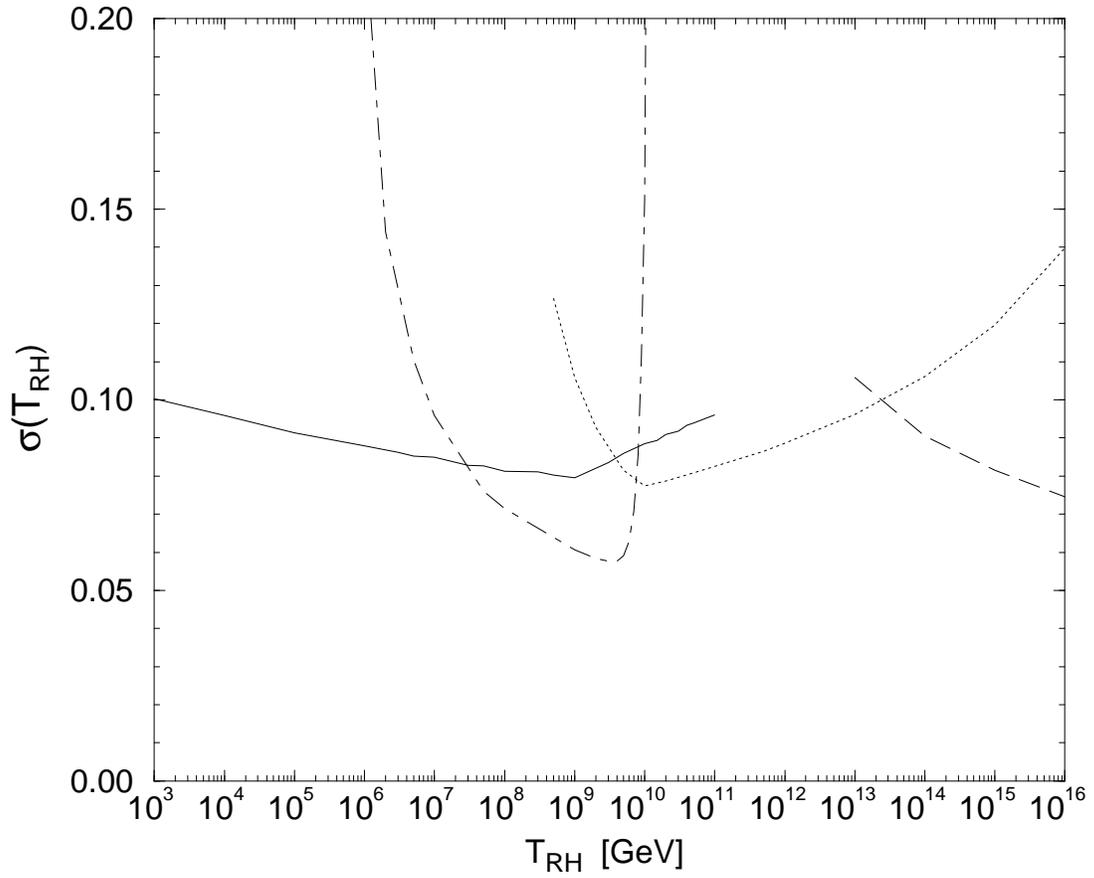}
\caption{Same as Fig.~\ref{bound-fig}, except that we show
the upper bound on $\sigma(T_{RH})$ as a function of $T_{RH}$.
This bound is independent of the spectral index $n$.}
\label{sigma-fig}
\end{figure}

We have calculated the density of primordial black holes using the
the near critical collapse mass function that results in a spectrum
of PBH masses for a given horizon mass.  The normalization of the
PBH mass spectrum was determined using the COBE anisotropy data
that allowed us to set bounds on the spectral index $n$ as a function
on the reheat temperature.  We find that restricting the density of PBHs 
to be less than the critical density corresponds to the restriction that
the spectral index $n$ be less than about $1.45$ to $1.2$, throughout 
the range of reheating temperatures resulting after inflation, 
$10^{3}$ to $10^{16}$ GeV respectively.  
(The precise limits are shown in Fig.~\ref{bound-fig}.)
For a smaller range of reheating temperatures, between about 
$10^{7}$ to $10^{10}$ GeV, significant PBH evaporation occurs 
when the optical depth of the universe is less than one.
Hence, we found a slightly stronger bound on the spectral index
by restricting the cosmological PBH evaporation into photons 
to be less than the present-day observed diffuse gamma-ray flux.
Due to the extreme sensitivity of the PBH mass density to the spectral index, 
effects such as the indirect photon flux from PBH evaporation into 
quarks and gluons which fragment into pions or the formation of a 
QCD photosphere are completely negligible when calculating the bound 
on $n$.  We should also remark that slightly stronger bounds on $n$ for 
larger reheating temperatures $\gtrsim 10^{10}$ GeV are expected
from PBHs that decay during the epoch of nucleosynthesis.

\begin{figure}[!t]
\centering
\epsfxsize=6.4in
\hspace*{0in}
\epsffile{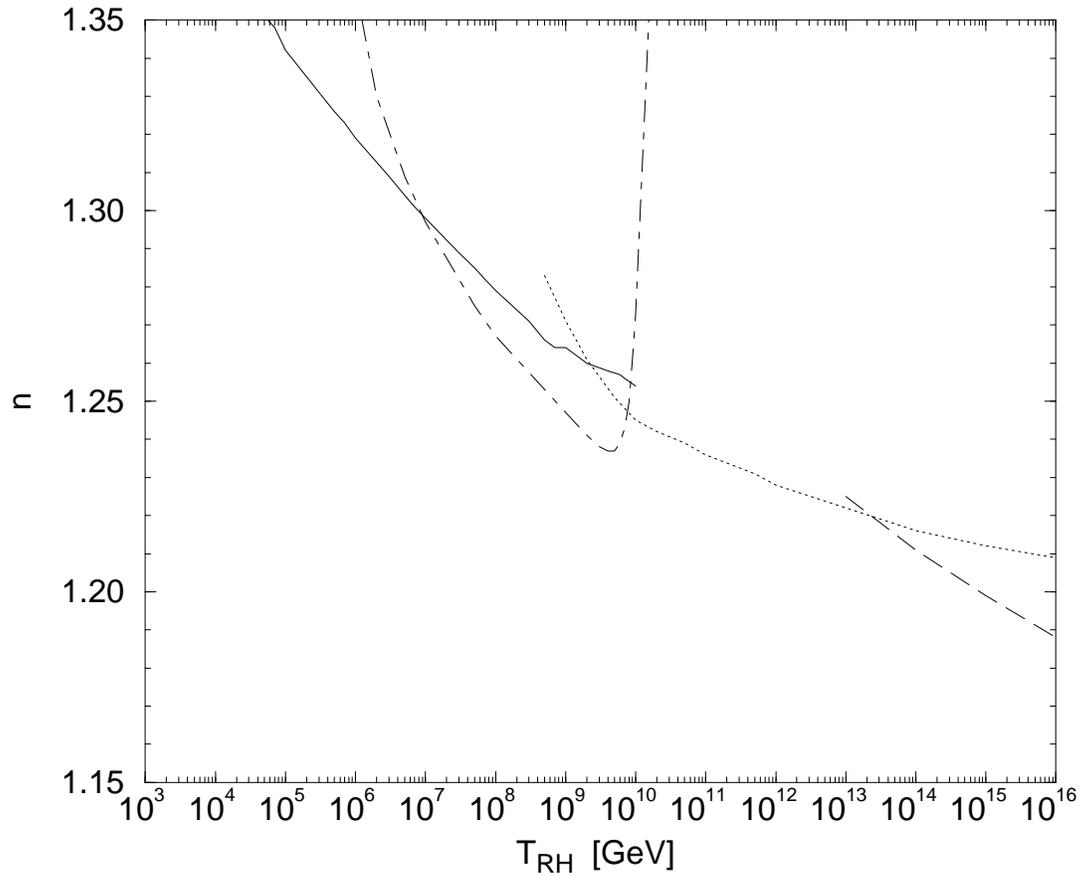}
\caption{The upper bound on $n$ as a function of $T_{RH}$ as in 
Fig.~\ref{bound-fig}, except that $\Omega_\Lambda = 0.7$.}
\label{bound-lambda-fig}
\end{figure}

If the universe is vacuum-energy dominated, there are corrections
to our bounds on $n$ that can be substantial.  We calculated these
corrections for a range of $\Omega_\Lambda$ and applied them to
our bounds on $n$ for the case of $\Omega_\Lambda = 0.7$.  The
improvement is apparent by contrasting Fig.~\ref{bound-fig} with
Fig.~\ref{bound-lambda-fig}.  Finally, we calculated bounds on the mass 
variance at reheating.  These bounds in principle could be used to 
constrain any given inflationary model, once the power spectrum 
is calculated.

\acknowledgments

This work was supported in part by the Department of Energy under
grant number DOE-ER-40682-143.  We thank Rich Holman and Jane MacGibbon 
for useful discussions.  We also thank Andrew Liddle useful 
discussions and comments on the manuscript.

{\tighten

}%end tighten (references)

\newpage

%%%%%%%%%%%%%%%%%%%%%%%%%%%%%%%%%%%%%%%%%%%%%%%%%%%%%%%%%%%%%%%%%%%%%
%%%%%%%%%%%%%%%%%%%%%%%%%%%%%%%%%%%%%%%%%%%%%%%%%%%%%%%%%%%%%%%%%%%%%

\newpage

%%%%%%%%%%%%%%%%%%%%%%%%%%%%%%%%%%%%%%%%%%%%%%%%%%%%%%%%%%%%%%%%%%%%%
%%%%%%%%%%%%%%%%%%%%%%%%%%%%%%%%%%%%%%%%%%%%%%%%%%%%%%%%%%%%%%%%%%%%%

\newpage

%%%%%%%%%%%%%%%%%%%%%%%%%%%%%%%%%%%%%%%%%%%%%%%%%%%%%%%%%%%%%%%%%%%%%
%%%%%%%%%%%%%%%%%%%%%%%%%%%%%%%%%%%%%%%%%%%%%%%%%%%%%%%%%%%%%%%%%%%%%

\newpage

%%%%%%%%%%%%%%%%%%%%%%%%%%%%%%%%%%%%%%%%%%%%%%%%%%%%%%%%%%%%%%%%%%%%%
%%%%%%%%%%%%%%%%%%%%%%%%%%%%%%%%%%%%%%%%%%%%%%%%%%%%%%%%%%%%%%%%%%%%%

\newpage

%%%%%%%%%%%%%%%%%%%%%%%%%%%%%%%%%%%%%%%%%%%%%%%%%%%%%%%%%%%%%%%%%%%%%
%%%%%%%%%%%%%%%%%%%%%%%%%%%%%%%%%%%%%%%%%%%%%%%%%%%%%%%%%%%%%%%%%%%%%

\end{document}